# Novel Scintillation Material – ZnO Transparent Ceramics

P.A. Rodnyi, K.A. Chernenko, E.I. Gorokhova, S.S. Kozlovskii, V.M. Khanin, and I.V. Khodyuk

*Abstract* — ZnO-based scintillation ceramics for application in HENPA LENPA analyzers have been investigated. The following ceramic samples have been prepared: undoped ones (ZnO), an excess of zinc in stoichiometry (ZnO:Zn), doped with gallium (ZnO:Ga) and lithium (ZnO:Li). Optical transmission, x-ray excited emission, scintillation decay and pulse height spectra were measured and analyzed. Ceramics have reasonable transparency in visible range (up to 60% for 0.4 mm thickness) and energy resolution (14.9% at 662 keV $Cs^{137}$ gamma excitation). Undoped ZnO shows slow (1.6 μs) luminescence with maximum at 2.37 eV and light yield about 57% of CsI:Tl. ZnO:Ga ceramics show relatively low light yield with ultra fast decay time (1 ns). Lithium doped ceramics ZnO:Li have better decay time than undoped ZnO with fair light yield. ZnO:Li ceramics show good characteristics under alpha-particle excitation and can be applied for the neutral particle analyzers.

*Index Terms* — Scintillation counters, zinc compounds, luminescence, ceramics, fusion reactor instrumentation.

## I. INTRODUCTION

The project of building the International Thermonuclear Experimental Reactor (ITER) requires solving a large number of engineering and technical tasks. One of the tasks is to measure the hydrogen isotope composition in the plasma on the basis of detection of neutral particle fluxes. Another task is to determine the alpha-particle distribution function by means of the detection of the $He^0$ atoms energy spectra in the MeV range [1]. To solve these problems tandem analyzers of the neutral particles (HENPA LENPA) are used. The analyzers are parts of the fusion reactor diagnostic system. The main feature of these devices is a necessity to work for a long time without maintenance in the powerful background flux of neutrons and electrons. These conditions determine a special system of particle detection, aiming to reduce influence of the background.

Neutral particles, that leave the tokamak chamber, become ionized by falling through the stripping foil. After that the bunch of ions goes through magnetic and electron-optical systems. Electronic and magnetic systems separate the beam in the perpendicular directions. Deviation of the particles in the electric field is proportional to the ratio of charge to energy. In the magnetic field deviation is proportional to the ratio of charge to momentum. Thus, spatial-separation of the ion beam makes it possible to identify type of the particle and its energy by measuring the interaction point. A scintillation spectrometer with a multianod photomultiplier tube (PMT) is used as a position sensitive detector.

The noise suppression principle is based on the property of heavy particles to lose energy in matter faster than light ones. If the thickness of the scintillator is equal to the range of the heavy particles, the signal from light particles with the same energy will be much smaller and it becomes possible to produce amplitude separation of the signals. We can achieve discrimination of the electronic background by using a thin scintillator. All the above mentioned determine the following requirements for the scintillator: high light yield (LY) (for detecting 100 keV particles), short decay time (in order to avoid pileup of signals of the electronic background, which leads to signal amplitude increase), small decrease in scintillation efficiency (α/β ratio) when excited by α-particles, nonhygroscopicity, possibility to manufacture the scintillator in the form of a plate with thickness ≤ 20 μm.

The purpose of this work was to develop a scintillator that satisfies the requirements mentioned above. Single-crystal scintillators do not satisfy the last requirement, while common film scintillators used for the alpha-particle registration have relatively low light yield. An alternative to single-crystal scintillators are optical ceramics. Optical ceramics are polycrystalline materials obtained by pressing micro- or nano-sized powders at high temperatures. Ceramics have high mechanical properties, are not hydroscopic, and are generally transparent in the visible range.

In this paper optical transmission, x-ray excited emission, scintillation decay and pulse height spectra of ZnO-based optical scintillation ceramics are presented and discussed.

## II. EXPERIMENTAL

ZnO ceramics were manufactured by the method of uniaxial hot pressing in a high vacuum furnace [2] in the form of discs with a diameter of 24 mm and after polishing they had thickness of 1.0-1.5 mm. These ceramics were machined in the form of plates 1x1 $cm^2$ and 20 μm thickness on a quartz substrate, which can be used in the detectors described above. We prepared samples of undoped ZnO, doped with gallium (ZnO:Ga) or lithium (ZnO:Li) and with an excess of zinc in stoichiometry (ZnO:Zn). High-purity powders with a particle size of 90-700 nm were used to obtain ceramics of undoped ZnO, ZnO:Ga and ZnO:Li. The ceramic ZnO:Zn sample was

P.A. Rodnyi, K.A Chernenko, S.S. Kozlovskii, V.M. Khanin, and I.V. Khodyuk are with the Saint Petersburg State Polytechnical University, Polytechnicheskaya 29, St.Petersburg, 195251, Russia. (e-mail: khodyuk@gmail.com)

E.I. Gorokhova is with Research and technological institute of optical materials all-russia scientific center "S.I.Vavilov State Optical Institute" Babushkina 36, St.Petersburg,192171 Russia

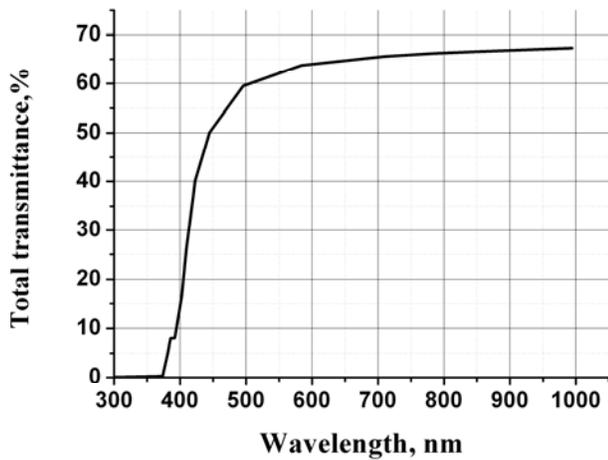

Figure 1. Total transmittance spectrum of undoped ZnO ceramics of 0.4 mm thickness.

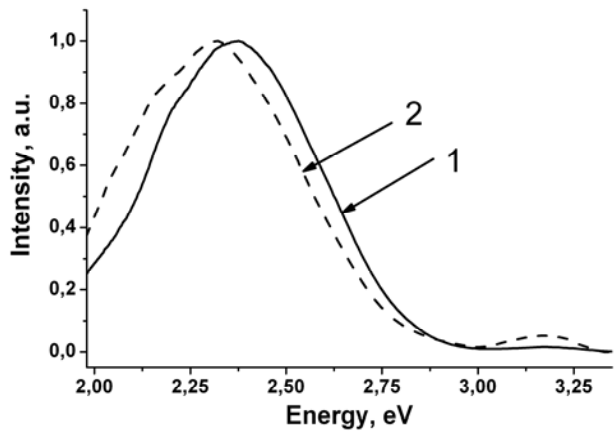

Fig. 2. Comparison of X-ray excited luminescence spectra of ZnO ceramics (1) and ZnO powder (2). Intensities are normalized.

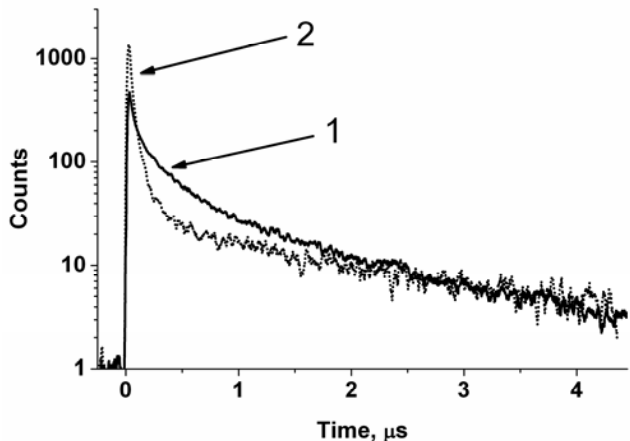

Fig. 3. Scintillation decay profiles of undoped ZnO (1) and ZnO:Li (2).

prepared from the electron-excited phosphor with a particle size of 3-10 μm.

The total transmittance spectra of the samples were measured using a Hitachi-330 spectrophotometer with an integral sphere of Ø 60 mm.

The X-ray excited luminescence (XRL) spectra were measured under steady state X-ray excitation (40 kV, 10 mA) with a FEU-106 PMT and a MDR-2 grating monochromator with 1200 grooves/mm. The spectra were corrected for the quantum efficiency of the PMT and for the monochromator transmission.

The decay time profiles were recorded with the experimental system described in detail in [3]. A pulsed X-ray source with 1 ns pulse duration operates at 30 kV and maximum current amplitude of 500 mA. The detection system consists of a FEU-71 PMT, operating in single photon counting mode plus related electronics. The system allows measurements with time resolution better than 100 ps. The XRL spectra and decay time curves were measured in the reflection mode: the angle between X-ray beam and the PMT was 90 deg.

The pulse height spectra of the ceramic scintillators were measured using a Hamamatsu R6231-100 PMT. The detected signal was fed to an Ortec 114 16k analog-to-digital converter via a preamplifier and an Ortec 672 spectroscopic amplifier. In order to improve the efficiency of photons collection upon the absorption of gamma radiation, the samples were covered with several layers of reflecting Teflon tape. Alpha-particle excited energy spectra were measured without covering.

III. EXPERIMENTAL RESULTS AND DISCUSSIONS

Prepared ZnO ceramics were highly transparent in the visible range. It should be noted that manufacturing highly transparent ceramics from materials with wurtzite (hexagonal) crystal structure is a difficult task. The transparency of the ceramics strongly depends on the parameters of manufacturing, purity, grain size, morphology of the initial powders, and the presence of different dopants and impurities. In particular, the heterogeneity of the powder grain decreases transparency. Undoped ZnO ceramics with thickness 0.4 mm has total transmittance of 60% at 500 nm wavelength (Fig. 1). Undoped ZnO ceramics and ceramics doped with lithium have the best transparency compared to the other samples, the total transmittance of 1.0 mm thick ceramics was 45% at 520 nm.

Typical XRL spectra of undoped ZnO ceramics (curve 1) and the initial powder (curve 2) are shown in Fig. 2. For both spectra green luminescence, also known as deep band emission (DBE), dominates. It has a large width at half maximum (500 meV) and apparently consists of several unresolved bands. The luminescence mechanism of this band is not fully understood and, in general, the following two centers are proposed to be responsible for the luminescence: zinc vacancies [4] and oxygen vacancies [5]. The authors of [6] showed that the powder's emission band with maximum at 2.35 eV is due to zinc vacancies, and the band with maximum at 2.53 eV is caused by oxygen vacancies. The XRL spectrum of ceramics, shown in Fig. 2, has a maximum at 2.37 eV, thus, the main luminescence centers are vacancies of zinc. However, the band has a large full width at half maximum and deviates from a Gaussian shape that indicates that other centers are involved in the luminescence process.

The XRL intensity of the initial powder is much lower than that of obtained ceramics. The maximum of the spectrum is shifted to the long wavelength region (2.32 eV) in comparison to the maximum of the ceramics spectrum (Fig. 2, curve 2).

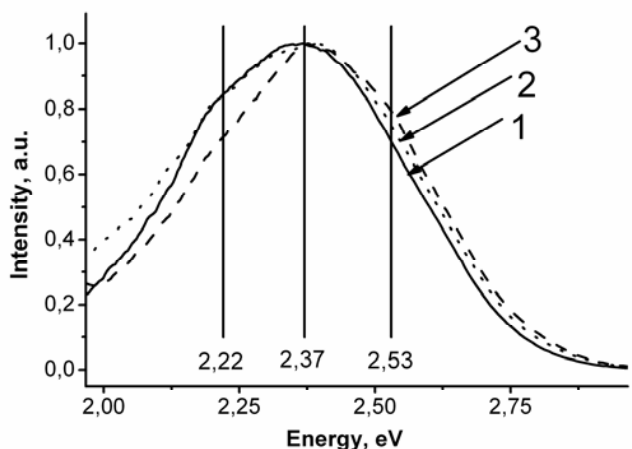

Fig. 4. X-ray excited luminescence spectra of ceramic ZnO (1), ZnO:Li(2), and ZnO:Zn (3). The intensities are normalized.

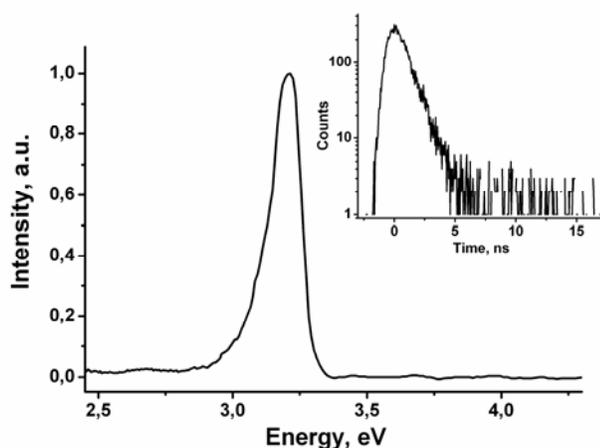

Fig. 5. X-ray excited luminescence spectra of ceramic ZnO:Ga$^{3+}$. Inset: scintillation decay profile of ZnO:Ga$^{3+}$.

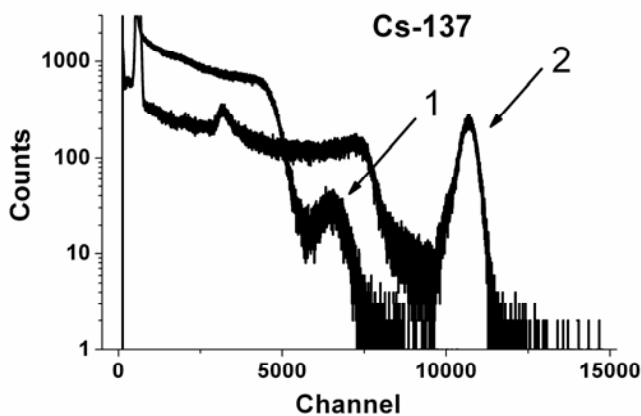

Fig 6. Pulse height spectra of $^{137}$Cs, obtained for ZnO ceramics (1) and CsI:Tl single crystalline (2) scintillators.

Undoped ZnO (Fig. 3, curve 1) has a long decay time (~1.6 μs). The hyperbolic shape of the decay curve indicates presence of the shallow traps. These traps can be associated with interstitial zinc (Zn$_i$) or other impurities [7].

The DBE also dominates in the XRL spectra of ZnO:Li and ZnO:Zn ceramics as shown in Fig. 4. It is considered that the oxygen vacancies (emission band peaking at 2.53 eV [6]) are the main luminescence centers in ceramics with an excess of zinc in stoichiometry. Higher intensity of the 2.53 eV band in the spectrum of ZnO:Zn ceramics can be seen in Fig. 4 (curve 3), compare to other ZnO-based ceramics. Also, the XRL spectrum of ZnO:Zn contains bands peaking at 2.37 eV (zinc vacancy) and at ~ 2.10 eV (unknown impurities or defects).

As shown in Fig. 4, the XRL spectrum of ceramic ZnO:Li has slightly higher relative intensity in the long wavelength region in comparison with undoped ceramic ZnO. Although the spectra of undoped ceramics and ceramics doped with lithium show little difference, their scintillation decay curves differ significantly (Fig. 3). Two main time components can be identified in the decay profile of ZnO:Li ceramics: fast (17 ns) and slow (~ 1 μs). The fast decay component contains about 60-80% of the total LY (depends on the dopant concentration).

Differences in the shape of the ZnO:Li spectrum (Fig. 3, curve 2) compared to the shape of the undoped ZnO (curve 1) are caused by presence of lithium ions, which form acceptor levels and neutralize free charges. According to [6, 8] lithium emission in ZnO powders is associated with band peaking at 2.17 eV. But as can be seen in Fig. 3 this band is not present at the XRL spectrum of the prepared ceramics.

The XRL spectra of ZnO ceramics doped with Ga$_2$O$_3$ have a significantly suppressed intensity of DBE. By varying dopant concentration and synthesis parameters it is possible to produce samples with only a short-wavelength emission band, with a maximum at 3.21 eV (Fig. 5) [9]. This band is associated with emission of the Wannier excitons [10]. Gallium ions replace zinc ions, as well as fill zinc vacancies, in the crystalline structure of the grains, which results in suppression of intensity of the DBE. The decay time constant of excitonic emission is about 1 ns (inset Fig. 5). The shape of the rise time of ZnO:Ga$^{3+}$ is determined by the duration of the X-ray excitation pulse. Unfortunately the LY of ZnO:Ga ceramics is only 2% of the LY of the undoped one.

Pulse height spectra of $^{137}$Cs, obtained with 0.4 mm thick ZnO ceramics and a standard 0.8 mm thick CsI(Tl) scintillation crystal, are shown in Fig. 6. To collect the slow component of the scintillation decay, the shaping time of the spectroscopic amplifier was set 10 μs. The LY of ZnO ceramics is about 57% of that of a CsI(Tl) single crystal at 662 keV gamma excitation. An energy resolution of 14.9% was measured for undoped ZnO ceramics under the conditions mentioned above. It is important to mention that no optical coupling was used between the scintillators and the photocathode of the PMT. By applying an optical coupling light collection efficiency and energy resolution can be improved.

ZnO ceramic samples with a thickness of 20 μm were investigated under alpha-particle excitation. Pulse height spectra of an $^{241}$Am alpha-source, obtained for 20 μm thick ceramic ZnO:Li and sputtered on a quartz plate CsI(Tl) scintillator of the same thickness are shown in Fig. 7. The energy of the alpha particles, emitted by the $^{241}$Am alpha-source was 5.5 MeV. Before absorption by the scintillator an alpha-particle passes through 20 mm of air. Part of the initial 5.5 MeV energy of the particle is lost so that when it reaches the scintillator surface it has energy about 3.5 MeV. According to our calculations the ranges of 3.5 MeV alphas in ZnO and CsI are about 8.61 µm and 15.4 µm, respectively The alpha-particle range is much shorter than the scintillator thickness so the energy of an alpha-particle is completely released in the scintillator. The energy resolution of the peak recorded with ZnO:Li (Fig. 7) ceramics is 62%. The LY of ZnO:Li is about 25% of that for CsI(Tl).

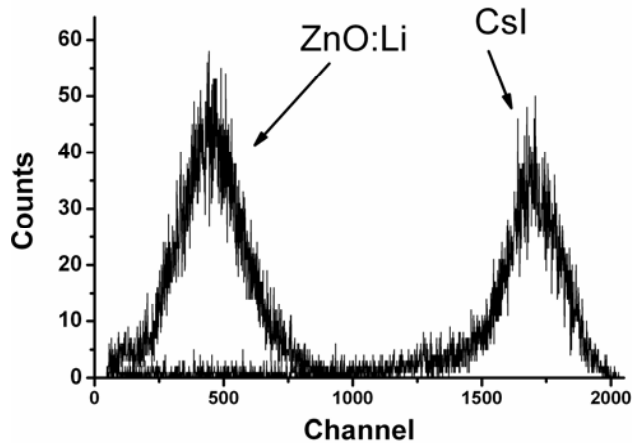

Fig. 7. Pulse height spectra of $^{241}$Am alpha-source, obtained for 20 µm thick ceramic ZnO:Li and sputtered on a quartz plate CsI(Tl) scintillator.

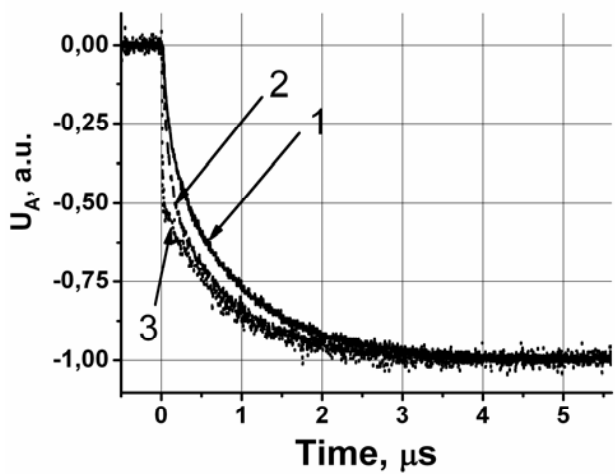

Fig.8. Time profile of the signal from the anode of the PMT under alpha-particle excitation for ZnO (1), ZnO:Li (2), and ZnO:Ga (3).

To estimate the scintillation decay time constants of the samples excited by alpha particles, the anode signal from the PMT was recorded by the oscilloscope (Fig. 8). The amplitude of the signal was normalized for different ceramics.

The curves shown in Fig. 8 can be considered as an integral of the decay time curve. A comparison of the scintillation decay curves shown in Fig. 3 and inset Fig. 5 with the shape of the signal obtained under alpha-excitation Fig. 8 shows an increase in the intensity of the slow luminescent component in case of alpha-excitation. This effect can be caused by peculiarities in the interaction of alpha-particles with solid materials: their short range and high ionization density.

## IV. CONCLUSION

The X-ray excited emission of $V_{Zn}$ centers with a maximum at 2.37 eV dominates in undoped ZnO ceramics. The LY of undoped ZnO ceramics is about 57% of that of CsI(Tl) with gamma excitation. The energy resolution of undoped ZnO ceramics is 14.9% at 662 keV gamma excitation. The average decay time is 1.6 µs.

Doping ZnO with lithium improves the scintillation characteristics of the ceramics. Scintillation decay time is reduced compared to undoped ZnO and the fast decay time component 17 ns contains 60-80% of the total emitted light.

ZnO doped with gallium does not show any slow component of the X-ray excited luminescence, but the LY is significantly quenched (~ 2% of the undoped ZnO). The main reason for the low light yield is the poor transparency of the ceramics at its own emission wavelength – selfabsorption.

A group of 20-µm thick samples fabricated and studied in this research showed that the scintillation properties of thin samples can be reproduced well. The performed experiments show that ZnO ceramics can be applied in the neutral particle analyzers.

Transparency in the visible range, high mechanical properties, low cost, and good match of the emission wavelength with a solid-state photodetector sensitivity (e.g., CCD matrix) make ZnO ceramics promising for use in X-ray computed tomography.

It should be noted that the emission wavelength of ZnO:Ga ceramics is in a good agreement with the spectral sensitivity of silicon photomultipliers (SiPM). ZnO:Ga has ultra fast decay time (1ns). At the same time SiPMs have very high time resolution. It seems to be very attractive to use SiPMs in conjunction with ZnO:Ga ceramics for applications where fast timing is essential.